\documentclass[twocolumn,twoside,slac_two]{revtex4}
\usepackage{graphicx}
\usepackage{fancyhdr}
\begin{document}

\title{On Equations for the Multi-quark Bound States in
Nambu--Jona-Lasinio Model}

%

\author{R.G. Jafarov}
\affiliation{Institute for Physical Problems, Baku State
University, AZ 1148, Baku, Azerbaijan}\email{jafarov@hotmail.com}
\author{V.E. Rochev}
\affiliation{Institute for High Energy Physics, 142280, Protvino,
Moscow Region, Russia}\email{rochev@ihep.ru}

\begin{abstract}
In present report we review some preliminary results of
investigation of higher orders of mean-field expansion for
Nambu--Jona-Lasinio model.  We discuss first results of
investigation of next-to-next-to-leading order of mean-field
expansion equations for four-quark and three-quark Green
functions. We have considered equations for Green functions of
Nambu--Jona-Lasinio model in mean-field expansion up to third
order.

\end{abstract}

\maketitle

\thispagestyle{fancy}


\section{INTRODUCTION AND SUMMARY}

Multi-particle equations are a traditional basis for a
quantum-field description of bound states in particle physics. A
well-known example of these equations is the two-particle
Bethe-Salpeter equation (BSE) for the two-particle amplitude and
for the two-particle bound state \cite{bs}.The multi-particle
(three or more particle) generalizations of the BSE have been also
studied. A straightforward generalization of two-particle BSE has
been intensively studied in sixties-seventies of last century. A
best exposition of these studies can be found in the work of Huang
and Weldon \cite{hw}.\\
\indent An essential imperfection of the original Bethe-Salpeter
approach to multi-particle equations was a full disconnection of
the approach with the field-theoretical equations for Green
functions (which are known as Schwinger-Dayson equations (SDE)).
This imperfection has been eliminated by Dahmen and Jona-Lasinio,
which had included the BSE to the field-theoretical Lagrangian
formalism with the consideration of functional Legendre
transformation with respect to bilocal source of fields
\cite{djl}. Then this approach has been generalized for
multi-particle equations with consideration of Legendre
transformation with respect to multi-local
sources \cite{r}.\\
\indent However, these theoretical constructions had not solved
the principal dynamical problem of quantitative description of
real bound states (nucleons, mesons etc.). A solution of BSE-type
equations has been founded as a very complicated mathematical tool
even for simple dynamical model.  There is a main reason of a
comparatively small popularity of the method of multi-particle
BSE-type equations among the theorists. Much more popular approach
to the problem of hadronization in QCD is based on the 't Hooft's
conjecture that QCD can be regarded as an effective theory of
mesons and glueballs \cite{th}. Subsequently, it was shown by
Witten that the baryons could be viewed as the solitons of the
meson theory \cite{w}. Futher development of these ideas has been
successful and has leaded to the prediction of pentaquark states
in baryon spectrum \cite{dpp}.\\
\indent Nevertheless, the investigations of multi-quark equations
are of significant interest due to the much less model assumptions
in this approach in comparison with the chiral-soliton models. The
solutions of multi-quark equations will provide us almost
exhaustive information about the structure of hadrons. There is
basic motivation of present work.\\
\indent We shall investigate Nambu--Jona-Lasinio (NJL) model with
quark content which is one of the most successful effective models
of QCD in the non-perturbative region (for review see \cite{vw},
\cite{k}). In overwhelming majority of the investigations, the NJL
model has been considered in the mean-field approximation or in
the leading order of $1/n_c$- expansion. However, a number of
perspective physical applications of NJL model is connected with
multi-quark functions (for example: meson decays, pion-pion
scattering, baryons, pentaquarks etc.). These multi-quark
functions arise in higher orders of mean-field expansion(MFE) for
NJL model. To formulate MFE we have used an iteration scheme of
solution of SDE with fermion bilocal source \cite{R}.\\
\indent We have considered equations for Green functions of NJL
model in MFE up to third order. The leading approximation and
next-to-leading order (NLO) of MFE maintain equations for the
quark propagator and the two-quark function and also the NLO
correction to the quark propagator. The next-to-next-to-leading
order (NNLO) of MFE maintains the equations for four-quark and
three-quark functions, and next-to-next-to-next-to-leading
order(NNNLO) of MFE maintains the equations for six-quark and
five-quark functions.

\section{MEAN-FIELD EXPANSION IN BILOCAL-SOURCE FORMALISM. LEADING ORDER
AND NLO}

\indent We consider NJL model with the Lagrangian

$$
{\cal L}=\bar \psi i\hat \partial\psi+\frac{g}{2}
\biggl[(\bar\psi\psi)^2+(\bar\psi i\gamma_5 \tau^a\psi)^2\biggr].
\label{LSU2}
$$

The Lagrangian is invariant under transformations of chiral group
$SU_V(2)\times SU_A(2)$, which correspond to u-d quark sector.
\indent A generating functional of Green functions (vacuum
expectation values of $T$-products of fields) can be represented
as the functional integral with bilocal source:
$$
G(\eta)  =\int D(\psi,\bar\psi)\exp i\Big\{\int dx{\cal L}$$
$$ -\int dx dy \bar\psi(y)\eta(y,x)\psi(x)\Big\}. \label{G}
$$
Here $\eta(y,x)$ is the bilocal source of the quark field.

\indent The n-th functional derivative of $G$   over source $\eta$
is the n-particle  (2n-point) Green function:
$$
\frac{\delta^n
G}{\delta\eta(y_1,x_1)\cdots\delta\eta(y_n,x_n)}\bigg\vert_{\eta=0}$$
$$= i^n<0 T
\Big\{\psi(x_1)\bar\psi(y_1)\cdots\psi(x_n)\bar\psi(y_n)\Big\}\mid0>$$
$$\equiv S_n\left(
\begin{array}{cc}x_1&y_1\\\cdots&\cdots\\x_n&y_n\end{array} \right).
\label{VEV}$$ Translational invariance of the
functional-integration measure gives us the
functional-differential SDE for the generating functional $G$ :

$$\delta(x-y)G + i\hat\partial_x\frac{\delta G}{\delta\eta(y,x)}
+ig\Big\{\frac{\delta}{\delta\eta(y,x)}tr {\biggl[}\frac{\delta
G}{\delta\eta(x,x)}{\biggl]}$$
$$-\gamma_5\tau^a\frac{\delta}{\delta\eta(y,x)}tr{\biggl[}\gamma_5\tau^a\frac{\delta
G}{\delta\eta(x,x)}{\biggl]}\Big\}$$

$$=\int dx_1\eta(x,x_1) \frac{\delta G}{\delta\eta(y,x_1)}.$$

\indent We  shall solve this equation employing the method which
proposed in work \cite{jr}. A leading approximation is the
functional
$$
G^{(0)}= \exp\Big\{\mbox{Tr}\,\Big(S^{(0)}\ast\eta\Big)\Big\}
\label{G^0}.
$$
The leading approximation generates the linear iteration scheme:
$$
G = G^{(0)} + G^{(1)} + \cdots + G^{(n)} + \cdots,
$$

Functional $G^{(n)}$ is
$$
G^{(n)}= P^{(n)}G^{(0)},
$$
where  $P^{(n)}$ is a polynomial of  $2n$ -th order over the
bilocal source $\eta$ .

\indent The unique connected Green function of the leading
approximation is the quark propagator. Other connected Green
functions  appear in the following iteration steps. The quark
propagator in the chiral limit is
$$
S^{(0)}= (m-\hat p)^{-1},
$$
where $m$ is the dynamical quark mass, which is a solution of gap
equation.

\indent A first-order functional is
$$ G^{(1)}=
\biggl\{\frac{1}{2}\mbox{Tr}\Big( S^{(1)}_2\ast\eta^2\Big)
+\mbox{Tr}\Big(S^{(1)}\ast\eta\Big)\biggr\}G^{(0)}.$$

The iteration-scheme equations give us the equation for
two-particle function $S^{(1)}_2$:

$$S_2^{(1)}\left( \begin{array}{cc} x&y\\x'&y'\end{array} \right)=-S_0(x-y')
S_0(x'-y)\label{S2}$$
$$+ig\int dx_1\Big\{(S_0(x-x_1)
S_0(x_1-y)) tr\biggl[ S_2^{(1)}\left( \begin{array}{cc}
x_1&x_1\\x'&y'\end{array} \right)\biggr]$$
\begin{equation}
- (S_0(x-x_1)\gamma_5\tau^a{\bf \tau}^{a_1} S_0(x_1-y)) tr\biggr[
\gamma_5\tau^a{\bf \tau}^{a_1}S_2^{(1)}\left(
\begin{array}{cc} x_1&x_1\\x'&y'\end{array} \right)\biggr]\Big\}
\end{equation}

and the NLO correction  to quark propagator $S^{(1)}$:

$$S^{(1)}(x-y)= ig\int dx_1S^{(0)}(x-x_1) \Big\{S_2^{(1)}\left(
\begin{array}{cc} x_1&y\\x_1&x_1\end{array} \right)$$
$$-\gamma_5 \tau^aS_2^{(1)}\left( \begin{array}{cc}
x_1&y\\x_1&x_1\end{array} \right)\gamma_5\tau^a\Big\} \label{S1}$$
$$+ ig\int dx_1 S^{(0)}(x-x_1)S^{(0)}(x_1-y)tr S^{(1)}(0).$$

The graphical representations of these equations see on Figs. 1
and 2., where the graphical notations of Fig. 3 are used.

\begin{figure}
\includegraphics[width=65mm]{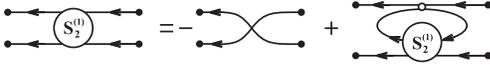}
\caption{The equation for NLO two-quark function.}
\end{figure}

\begin{figure}
\includegraphics[width=65mm]{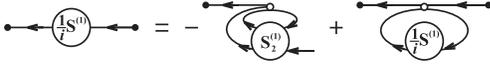}
\caption{The equation for NLO correction to quark propagator.}
\end{figure}

\begin{figure}
\includegraphics[width=65mm]{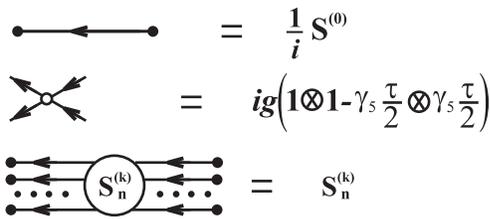}
\caption{Diagram rules.}
\end{figure}

\indent These equations can be easily solved, and the solutions
contain singlet scalar quark-antiquark bound state with mass $2m$
(sigma-meson) and massless (in the chiral limit) pseudoscalar
bound states (pions). To describe the solution of the NLO equation
for two-quark function and for future purposes we introduce  the
composite meson propagators by following way:

\indent a) Let us define scalar-scalar function

\begin{equation}
S_\sigma(x-x')\equiv tr\biggl[S_2^{(1)}\left( \begin{array}{cc}
x&x\\x'&x'\end{array}\right)\biggr]\sim
<\bar{\psi}\psi(x)\bar{\psi}\psi(x')>
\end{equation}
From the equation (1) for two-quark function we obtain (in
momentum space)
\begin{equation}
S_\sigma(p)=\frac{1}{ig}(1-i\Delta_\sigma(p))
\end{equation}

Here we define the following function, which we call
$\sigma-$meson propagator

\begin{equation}
\Delta_\sigma(p)=\frac{Z(p)}{4m^2-p^2},
\end{equation}

where $$Z_\sigma(p)=\frac{I_0(4m^2)}{I_0(p^2)}$$ and
$$I_0(p )=\int d \tilde q\frac{1}{(m^2-(p+q)^2)(m^2-q^2)}.$$

\indent b) Pseudoscalar-pseudoscalar function is defined as

$$S_\pi^{ab}(x-x')\equiv tr\biggl[S_2^{(1)}\left( \begin{array}{cc}
x&x\\x'&x'\end{array} \right) \gamma_5\frac{{\bf
\tau}^a}{2}\gamma_5 \frac{{\bf \tau}^b}{2}\biggr]$$
\begin{equation}\sim<\bar{\psi}\gamma_5\frac{{\bf \tau}^a}{2}\psi(x)
\bar{\psi}\gamma_5\frac{{\bf \tau}^b}{2}\psi(x')>
\end{equation}
From the equation (1) for two-quark function we obtain (in
momentum space):
\begin{equation}
S_\pi^{ab}(p)=-\frac{1}{ig}(\delta^{ab}-i\Delta_\pi^{ab}(p)).
\end{equation}

Here we define the pion propagator

\begin{equation}
\Delta_\pi^{ab}(p)=-\frac{\delta^{ab}Z(p)}{p^2},
\end{equation}

where $Z_\pi(p)=\frac{I_0(0)}{I_0(p^2)}$.

\section{NNLO OF MEAN-FIELD EXPANSION}

\indent NNLO(second-order)generating functional is

$$G^{\left( 2\right) }\left[ \eta \right] =\biggl\{\frac {1}{4!}\mbox{Tr}
\Big(S^{(2)}_4\ast\eta^4\Big)+ \frac {1}{3!}\mbox{Tr}
\Big(S^{(2)}_3\ast\eta^3\Big)$$
$$+\frac {1}{2}\mbox{Tr}
\Big(S^{(2)}_2\ast\eta^2\Big)+\mbox{Tr}\Big(S^{(2)}\ast\eta\Big)\biggr\}G^{(0)}.$$
\bigskip
The equations for four-quark and three-quark functions see on
Figs. 4 and 5.

\begin{figure}
\includegraphics[width=65mm]{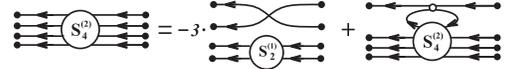}
\caption{The equation for NNLO four-quark function.}
\end{figure}

\begin{figure}
\includegraphics[width=65mm]{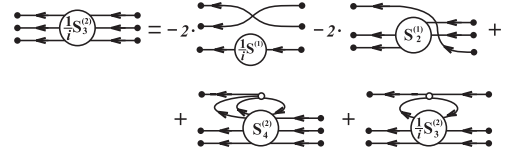}
\caption{The equation for NNLO three-quark function.}
\end{figure}

\indent The equations for the four-quark function $S^{(2)}_4$  and
for the three-quark functions  $S^{(2)}_3$ are new, and the
equations for two-particle function $S^{(2)}_2$   and propagator
$S^{(2)}$ have the same form as the corresponding NLO equation
except of the inhomogeneous terms. For NNLO equations these terms
contain four-quark function $S^{(2)}_4$ and three-quark
$S^{(2)}_3$function.

\indent The equation for the four-quark function has a simple
exact solution which is the product of NLO two-quark functions
(see Fig. 6). As it seen from this solution,  the
$\pi\pi-$scattering in NJL model is suppressed, i.e. in the NNLO
of MFE this scattering is absent, and it can be arise in the NNNLO
(third order) only.

\begin{figure}
\includegraphics[width=65mm]{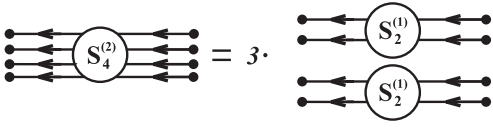}
\caption{The solution of equation for four-quark function.}
\end{figure}

\section{VERTEX $\sigma\pi\pi$}

\indent The existence of above exact solution for the four-quark
function gives us a possibility to obtain a closed equation for
the three-quark function. As a first step in an investigation of
this  equation we shall solve a problem of definition of
$\sigma\pi\pi-$vertex with composite sigma-meson and pions. Let us
introduce a function
$$W_{\sigma\pi\pi}^{ab}(xx'x'')
\equiv tr\biggl[S_3^{(2)}\left(
\begin{array}{ccc} x&x\\x'&x'\\x''&x''\end{array}
\right)\begin{array}{ccc} \end{array}
\gamma_5\frac{\tau^a}{2}\gamma_5 \frac{\tau^b}{2}\biggr]$$
$$\sim <\bar{\psi}\psi(x)\bar{\psi}\gamma_5\frac{\tau^a}{2}\psi(x')
\bar{\psi}\gamma_5\frac{\tau^b}{2}\psi(x'')>
$$

and define:

a) scalar vertex

$$V_\sigma(xx'x'')\equiv
tr\biggl[S_0(x-x')
S_2^{(1)}\left(\begin{array}{cc}x'&x\\x''&x''\end{array}\right)\biggr]$$
\begin{equation}
=2in_c\int dx_1v_S(xx'x_1)\Delta_\sigma(x_1-x'').
\end{equation}
Here: $v_S(xx'x'')=tr_\alpha [S_0(x-x')S_0(x'-x'')S_0(x''-x)]$ is
the triangle diagram.

b) pseudoscalar vertex

$$
V_\pi^{ab}(xx'x'')\equiv tr\biggl[S_0(x-x')\gamma_5 \frac{{\bf
\tau}^a}{2}S_2^{(1)}\left(
\begin{array}{cc}
x'&x\\x''&x''\end{array} \right)\gamma_5\frac{{\bf
\tau}^b}{2}\biggr]
$$
\begin{equation}
=2in_c\int dx_1v_P(xx'x_1)\Delta_\pi^{ab}(x_1-x'').
\end{equation}
Here:
$v_P(xx'x'')=tr_\alpha[S_0(x-x')\gamma_5S_0(x'-x'')\gamma_5S_0(x''-x))].$

With definitions (2)-(9) we obtain for vertex function $W^{ab}$
the following equation:

$$
W_{\sigma\pi\pi}^{ab}(xx'x'')=W^{ab}_{0}(xx'x'')
$$
$$
+2ign_c\int dx_1l_S(x-x_1)W_{\sigma\pi\pi}^{ab}(x_1x'x''),
$$

where $l_S(x)\equiv tr_{\alpha} [S_0(x)S_0(-x)]$ is the scalar
quark loop. Inhomogeneous term  $W^{ab}_{0}$ is:
$$
W^{ab}_{0}(xx'x'')= V_\pi^{ab}(xx'x'')+V_\pi^{ab}(xx''x')
$$
$$
+4ig\int dx_1 V_\pi^{a_1a}(xx_1x')S_\pi^{a_1b}(x_1-x'')
$$
$$
+4ig\int dx_1 V_\pi^{a_1b}(xx_1x'')S_\pi^{a_1a}(x_1-x')
$$
$$
-ig\int
dx_1(V_\sigma(xx_1x_1)-4V_\pi^{a_1a_1}(xx_1x_1))S_\pi^{ab}(x'-x'')
$$

Using definitions (2)-(9) we have:
$$
[W_0^{ab}(xx'x'')]^{con}
$$
$$
=-2n_c\int dx_1 dx_2
v_P(xx_1x_2)[\Delta_\pi^{a_1a}(x_2-x')\Delta_\pi^{a_1b}(x_1-x'')
$$
$$
+\Delta_\pi^{a_1ab}(x_2-x'')\Delta_\pi^{a_1a}(x_1-x')].
$$

The equation for  $W^{ab}$ can be easy solved in the momentum
space and and a solution is
$$
W_{\sigma\pi\pi}^{ab}(pp'p'')=i\Delta_\sigma(p)W_0^{ab}(pp'p'')
$$

where $p$ is $\sigma-$mesons momenta, and $p'$, $p''$ are pion
momenta: $p=p'+p''$. The connected part of $W^{ab}$ is an decay
amplitude  $\sigma\rightarrow\pi\pi$. It is has a following form:
$$
[W^{ab}_{\sigma\pi\pi}(pp'p'')]^{con}
$$

\begin{equation}
=\frac{2n_c}{i}\Delta_\sigma(p) [v_P(pp'p'')+v_P(pp''p')]
\Delta_\pi^{aa_1}(p')\Delta_\pi^{a_1b}(p''),
\end{equation}
(See also Fig.7.).
\begin{figure}
\includegraphics[width=65mm]{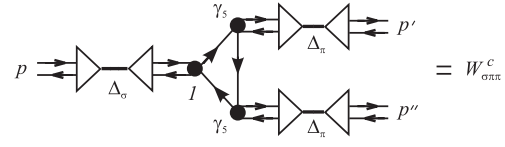}
\caption{The connected part of sigma-pion-pion-vertex.}
\end{figure}

    \section{NNNLO OF MEAN-FIELD EXPANSION}

\indent The third-order generating functional is
$$
G^{\left( 3\right) }\left[ \eta \right] =\biggl\{\frac
{1}{6!}\mbox{Tr} \Big(S^{(3)}_6\ast\eta^6\Big)+ \frac
{1}{5!}\mbox{Tr} \Big(S^{(3)}_5\ast\eta^5\Big)
$$
$$
+\frac {1}{4!}\mbox{Tr}\Big(S^{(3)}_4\ast\eta^4\Big)+ +\frac
{1}{3!}\mbox{Tr}\Big(S^{(3)}_3\ast\eta^3\Big)
$$
$$
+\frac {1}{2}\mbox{Tr}\Big(S^{(3)}_2\ast\eta^2\Big)+ \mbox{Tr}
\Big(S^{(3)}\ast\eta\Big)\biggr\}G^{(0)}.
$$

The graphical representations of equations for six-quark functions
and for five-quark function see on Figs. 8 and 9. The equations
for the six-quark function and for the five-quark function in our
iteration scheme are new, and the equations for four-quark
function $S^{(3)}_4$,three-quark function  $S^{(3)}_3$, two-quark
function $S^{(3)}_2$ and quark propagator $S^{(3)}$  have the same
form as the second-order equations  except of the inhomogeneous
term, which contains the six-quark function and the five-quark
function.

\begin{figure}
\includegraphics[width=65mm]{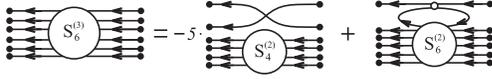}
\caption{The equation for six-quark function.}
\end{figure}

\begin{figure}
\includegraphics[width=65mm]{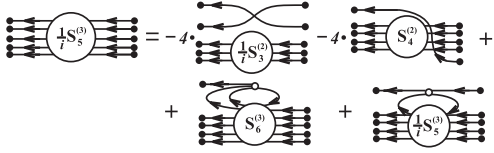}
\caption{The equation for five-quark function.}
\end{figure}

\begin{acknowledgments}
One of the authors (R.G.J.) would like to express his sincere
gratitude to the IPM and Organizing Committee for the invitation
and warm hospitality.
\end{acknowledgments}

\bigskip 

\end{document}